\documentclass[english,graybox]{svmult}
\usepackage[T1]{fontenc}
\usepackage[latin9]{inputenc}
\usepackage{units}
\usepackage{url}
\usepackage{amsmath}
\usepackage{undertilde}
\usepackage{graphicx}

\makeatletter

\providecommand{\tabularnewline}{\\}

\newenvironment{lyxlist}[1]
{\begin{list}{}
{\settowidth{\labelwidth}{#1}
 \setlength{\leftmargin}{\labelwidth}
 \addtolength{\leftmargin}{\labelsep}
 }}
{\end{list}}

\usepackage{units}
\usepackage{babel}
\newcommand{\be}{\begin{equation*}}
\newcommand{\ee}{\end{equation*}}

\makeatother

\usepackage{babel}
\begin{document}

\title*{Some Statistical Problems with High Dimensional Financial data}

\author{Arnab Chakrabarti and Rituparna Sen}

\institute{Arnab Chakrabarti \at Indian Statistical Institute \email{arnab@isichennai.res.in}
\and Rituparna Sen \at Indian Statistical Institute, Chennai \email{rsen@isichennai.res.in}}
\maketitle

\abstract{For high dimensional data some of the standard statistical techniques
do not work well. So modification or further development of statistical
methods are necessary. In this paper we explore these modifications.
We start with important problem of estimating high dimensional covariance
matrix. Then we explore some of the important statistical techniques
such as high dimensional regression, principal component analysis,
multiple testing problems and classification. We describe some of
the fast algorithms that can be readily applied in practice. }

\section{Introduction}

\label{sec:1} A high degree of interdependence among modern financial
systems, such as firms or banks, is captured through modeling by a
network $G(V,E)$, where each node in $V$ represents a financial
institution and an edge in $E$ stands for dependence between two
such institutions. The edges are determined by calculating the correlation
coefficient between asset prices of pairs of financial institutions.
If the sample pairwise correlation coefficient is greater than some
predefined threshold then an edge is formed between corresponding
nodes. This network model can be useful to answer important questions
on the financial market, such as determining clusters or sectors in
the market, uncovering possibility of portfolio diversification or
investigating the degree distribution \cite{boginski2005}, \cite{vandewalle2001}.
See figure 1 for illustration of one such network. Using correlation
coefficients to construct the economic or financial network has a
serious drawback. If one is interested in direct dependence of two
financial institutions the high observed correlation may be due to
the effect of other institutions. Therefore a more appropriate measure
to investigate direct dependence is partial correlation. Correlation
and partial correlation coefficients are related to covariance and
inverse covariance matrix respectively. Therefore in order to have
meaningful inference on the complex system of financial network, estimation
of covariance matrix accurately is of utmost importance. In this paper
we investigate how inference based on covariance matrix for high dimensional
data can be problematic and how to solve the problem. 

\begin{figure}
\label{fig1} \includegraphics[width=1\textwidth]{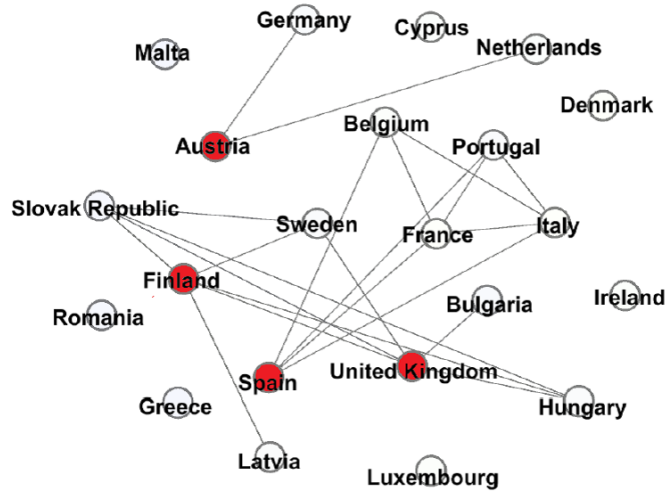}\caption{Network topology of European economies in post-euro period as described
in \cite{papadimitriou2014european}. }
\end{figure}

The rest of the paper is organized as follows. In section 2 we discuss
the distribution of eigenvalues of covariance matrix. In section 3
the problem and possible solution of covariance matrix estimation
is discussed. Section 4 deals with estimation of precision matrix.
Section 5 and 6 deals with multiple testing procedure and high dimensional
regression problem respectively. We discuss high dimensional principal
component analysis and several classification algorithms in section
7 and 8. 

\section{Distribution of Eigenvalues}

\subsection{Eigenvalues of Covariance Matrix}

\begin{figure}
\caption{Plot of true (dotted line) and sample (solid line) eigenvalues.}
\label{fig2}\includegraphics[width=0.95\textwidth]{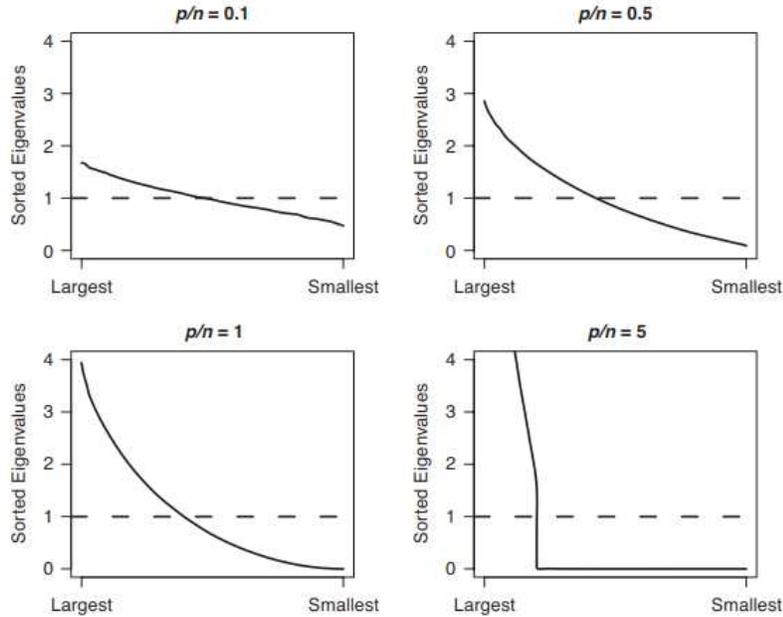} 
\end{figure}

In multivariate statistical theory, the sample covariance matrix is
the most common and undisputed estimator because it is unbiased and
has good large sample properties with growing number of observations
when number of variables is fixed. But if the ratio of the number
of variables ($p$) and the number of observations ($n$) is large,
then sample covariance does not behave as expected. It can be showen
that if $p$ grows at the same rate as $n$ i.e. $\nicefrac{p}{n}\rightarrow y\!>0)$
the sample covariance matrix becomes inconsistent and therefore can
not be relied upon \cite{stein1956}. In Figure \ref{fig2} eigenvalues
of population covariance matrix and sample covariance matrix are plotted
for different values of $p$ and $n$ where the population covariance
matrix is the identity matrix. It is evident that the true and sample
spectra differ a lot as the ratio $\nicefrac{p}{n}$ grows. So for
high dimensional data ($\nicefrac{p}{n}\rightarrow y>0$) there is
a need to find an improved estimator. 

Even though the sample eigenvalues are not consistent anymore, the
limiting distribution of eigenvalues of the sample covariance matrix
and the connection it has with the limiting eigenvalue distribution
of population covariance matrix are of importance. Determining the
limiting spectral distribution can be very useful to test the underlying
assumption of the model. In this section we will very briefly discuss
some results of random matrix theory that answers this kind of questions.
Throughout we will denote the ratio $p/n$ as $y_{n}$. 

\subsection{Marchenko Pastur Law and Tracy Widom Law}

Suppose that $\{x_{ij}\}$ are iid Gaussian variables with variance
$\sigma^{2}$. If $\nicefrac{p}{n}\rightarrow y>0$ then the empirical
spectral distribution (distribution function of eigenvalues) of sample
covariance matrix $S_{n}$converges almost surely to the distribution
F with the density 
\begin{align*}
f(x) & =\frac{1}{2\pi\sigma^{2}yx}\sqrt{(b-x)(x-a)}I(a\leq x\leq b)\\
\end{align*}
if $y<1$, where $a=a(y)=\sigma^{2}(1-\sqrt{y})^{2}$ and $b=b(y)=\sigma^{2}(1+\sqrt{y})^{2}$.
If $y$ $>1$ It will take additional positive mass $1-\frac{1}{y}$
at 0. 

$\sigma$ is called the scale parameter. The distribution is known
as Marchenko-Pastur distribution. 

If $\nicefrac{p}{n}\rightarrow0$ then empirical spectral distribution
of $W_{n}=\sqrt{\frac{n}{p}(S_{n}-\sigma^{2}I})$ converges almost
surely to the \textit{semicircle law} with density: 
\[
f(x)=\frac{1}{2\pi\sigma^{2}}\sqrt{4\sigma^{2}-x^{2}}I(|x|\leq2\sigma)
\]
Although the sample eigenvalues are not consistent estimator, the
limiting spectral distribution is related to the population covariance
matrix in a particular way. 

Also if $p\rightarrow\infty$ and $n\rightarrow\infty$ such that
$\frac{p}{n}\rightarrow y>0$, then $\frac{\lambda_{1}-\mu_{np}}{\sigma_{np}}\stackrel{\mathcal{L}}{\rightarrow}W_{1}$where
$\lambda_{1}$ is the largest eigenvalue of sample covariance $\mu_{np}=(\sqrt{n}+\sqrt{p})^{2}$
and $\sigma_{np}=(\sqrt{n}+\sqrt{p})(\frac{1}{\sqrt{n}}+\frac{1}{\sqrt{p}})^{\frac{1}{3}}$
and $W_{1}$ is Tracy-Widom Law distribution. 

\section{Covariance Matrix Estimator}

\subsection{Stein's Approach}

We see from figure \ref{fig2} that the sample eigenvalues can differ
a lot from the population eigenvalues. Thus shrinking the eigenvalues
to a central value is a reasonable approach to take. Such an estimator
was proposed by Stein \cite{stein1956inadmissibility} and takes the
following form: 
\[
\hat{\Sigma}=\hat{\Sigma}(S)=P\psi(\Lambda)P'
\]
where $\Lambda=diag(\lambda_{1},\lambda_{2},...,\lambda_{p})$ and
$\psi(\Lambda)$ is also a diagonal matrix. If $\psi(\lambda_{i})=\lambda_{i}\forall i$
then $\hat{\Sigma}$ is the usual estimator $S$. In this approach
the eigen vectors are kept as it is, but the eigenvalues are shrink
towards a central value. As the eigen vectors are not altered or regularized
this estimator is called rotation equivariant covariance estimator.
To come up with a choice of $\psi$, we can use entropy loss function
\[
L=\mathrm{tr}(\hat{\Sigma}\Sigma^{-1})-\mathrm{log}(\hat{\Sigma}\Sigma^{-1})-p
\]
or Frobeneous loss function 
\[
L_{2}=\mathrm{tr}(\hat{\Sigma}\Sigma^{-1}-I)^{2}.
\]
Under entropy risk ($=E_{\Sigma}(L)$), we have $\psi(\lambda_{i})=\frac{\lambda_{i}n}{\alpha_{i}}$
where 
\[
\alpha_{i}=(n-p+1+2\lambda_{i}\Sigma_{i\neq j}\frac{1}{\lambda_{i}-\lambda_{j}}).
\]

\noindent The only problem with this estimator is that some of the
essential properties of eigenvalues, like monotonicity and nonnegativity,
are not guaranteed. Some modifications can be adopted in order to
force the estimator to satisfy those conditions (see \cite{Lin} and
\cite{naul2016}). An algorithm was proposed to avoid such undesirable
situations by pooling adjacent estimators together in \cite{Stein1975}.
In this algorithm first the negative $\alpha_{i}$'s are pooled together
with previous values until it becomes positive and then to keep the
monotonicity intact, the estimates ($\psi's$) are pooled together
pairwise. 

\subsection{Ledoit-Wolf type Estimator}

As an alternative to the above mentioned method, the empirical Bayes
estimator can also be used to shrink the eigenvalues of sample covariance
matrix. \cite{haff1980} proposed to estimate $\Sigma$ by 
\[
\hat{\Sigma}=\frac{np-2n-2}{n^{2}p}\utilde{\alpha}I+\frac{n}{n+1}S,
\]
where $\utilde{\alpha}=(\mathrm{det}(S))^{\nicefrac{1}{p}}$. This
estimator is a linear combination of $S$ and $I$ which is reasonable
because although $S$ is unbiased, it is highly unstable for high
dimensional data and $\alpha I$ has very little variability with
possibly high bias. Therefore a more general form of estimator would
be 
\[
\hat{\Sigma}=\alpha_{1}T+\alpha_{2}S
\]
where $T$ is a positive definite matrix and $\alpha_{1}$ (shrinkage
intensity parameter), $\alpha_{2}$ can be determined by minimising
the loss function. For example \cite{ledoit2004} used 
\[
L(\hat{\Sigma},\Sigma)=\frac{1}{p}\mathrm{tr}(\hat{\Sigma}-\Sigma)^{2}
\]
to obtain a consistent estimator with $T=I$. A trade off between
bias and variance is achieved through the value of shrinkage parameter.
In figure 3, bias, variance and MSE are plotted against shrinkage
parameter. The optimum value of shrinkage intensity is that for which
MSE is minimum. 

It can be shown that if there exists $k_{1}$ and $k_{2}$ independent
of $n$ such that $\nicefrac{p}{n}\leq k_{1}$ and $\frac{1}{p}\sum_{i=1}^{p}\mathrm{E}[Y_{i}]^{8}\leq k_{2}$
where $Y_{i}$ is the $i$-th element of any row of the matrix of
principal components of $X$ and if 
\[
\lim_{n\rightarrow\infty}\frac{p^{2}}{n^{2}}\times\frac{\sum_{(i,j,k,l)\in Q_{n}}(\mathrm{Cov}[Y_{i}Y_{j},Y_{k}Y_{l}])^{2}}{\mathrm{cardinality}(Q_{n})}=0
\]
where $Q_{n}$ denotes the set of all quadruples made of four distinct
integers between $1$ and $p$, then the following estimator $S_{n}^{*}$
(a convex combination of $I$ and $S$) is consistent for $\Sigma$,
see \cite{ledoit2004}: 
\[
S_{n}^{*}=\frac{b_{n}^{2}}{d_{n}^{2}}m_{n}I_{n}+\frac{d_{n}^{2}-b_{n}^{2}}{d_{n}^{2}}S_{n}
\]
where $X_{k.}$ is the $k$th row of $X$ and 
\begin{align*}
m_{n} & =\frac{1}{p}\mathrm{tr}(S_{n}^{'}I_{n})\\
d_{n}^{2} & =\|S_{n}-m_{n}I_{n}\|^{2}\\
b_{n}^{2} & =\mathrm{min}(d_{n}^{2},\frac{1}{n^{2}}\sum_{k=1}^{n}\|X_{k.}^{'}X_{k.}-S_{n}\|^{2}).
\end{align*}

\noindent The first condition clearly deals with the interplay between
sample size, dimension and moments whereas the second one deals with
the dependence structure. For $\nicefrac{p}{n}\rightarrow0$ the last
condition for dependence structure can be trivially verified as a
consequence of the assumption on moments. This estimate is also computationally
easy to work with. In fact as $S$ is still unbiased estimator one
possible way to reduce the variance is to use bootstrap-dependent
techniques like bagging. But that is far more computationally demanding
compared to this method.

With the additional assumption that $\mathrm{var}(\frac{\sum_{i=1}^{p}Y_{i}^{2}}{p})$
is bounded as $n\rightarrow\infty$, \cite{ledoit2004} showed that
\[
lim_{n\rightarrow\infty}[E||S_{n}-\Sigma_{n}||^{2}-\frac{p}{n}(m_{n}^{2}+var(\frac{\sum_{i=1}^{p}Y_{i}^{2}}{p}))]=0
\]
This result implies that expected loss of sample covariance matrix,
although bounded, does not usually vanish. Therefore consistency of
usual sample covariance matrix is achieved only when $\frac{p}{n}\rightarrow0$
or $m_{n}^{2}+\mathrm{var}(\frac{\sum_{i=1}^{p}Y_{i}^{2}}{p})\rightarrow0.$
In the latter case most of the random variables are asymptotically
degenerate. The difference between these two cases is that in the
first, the number of variables is very less compared to $n$ and in
the latter $O(n)$ degenerate variables are augmented with the earlier
lot. Both of these essentially indicate sparsity.

\begin{figure}
\centering \includegraphics{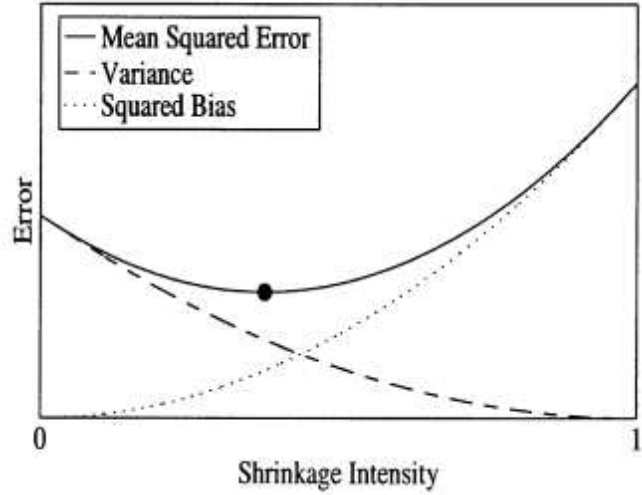}\caption{Plot for Error vs shirnkage intensity \cite{ledoit2004}}
\label{fig3} 
\end{figure}

A more general target matrix $T$ can be used instead $I$. For example,
under Gaussian distribution, if $T=\nicefrac{\mathrm{tr}(S)}{p}I$,
$\alpha_{1}=\lambda$ (intensity parameter) and $\alpha_{2}=1-\lambda$,
then optimal shrinkage intensity is 
\[
min(\frac{\sum_{i=1}^{p}||x_{i}x_{i}'-S||_{F}^{2}}{n^{2}[tr(S^{2})-tr^{2}(S)/p]},1)
\]
 which implies that the shrinkage estimator is a function of the sufficient
statistics $S$ and therefore can be further improved upon by using
Rao-Blackwell theorem \cite{chen2009shrinkage}. The resulting estimator
becomes $\lambda_{RBLW}T+(1-\lambda_{RBLW})S$ where 
\[
\lambda_{RBLW}=\frac{\frac{n-2}{n}tr(S)+tr^{2}(S)}{(n+2)[tr(S^{2})-\frac{tr^{2}(S)}{p}]}
\]
 If we take $T=\mathrm{Diag}(S)$, that is, the diagonal elements
of $S$ then the optimal intensity that minimises $E[\|\hat{\Sigma}-\Sigma\|^{2}]$
can be estimated as 
\[
\frac{\frac{1}{n}(\hat{a}_{2}+p\hat{a}_{12})-\frac{2}{n}\hat{a}_{2}*}{\frac{n+1}{n}\hat{a}_{2}+\frac{p}{n}\hat{a}_{1}^{2}-\frac{n+2}{n}\hat{a}_{2}*}
\]
 where $\hat{a}_{1}=\nicefrac{1}{p}\mathrm{tr}(S)$, $\hat{a}_{2}=\frac{n^{2}}{(n-1)(n+2)}\frac{1}{p}[\mathrm{tr}S^{2}-\frac{1}{n}(\mathrm{tr}S)^{2}],$
$\hat{a}_{2}^{*}=\frac{n}{n+2}\mathrm{tr}(T^{2})/p$ as shown in\cite{fisher2011}.
\cite{schafer2005} chose the shrinkage parameter to be 
\[
\lambda*=\frac{\sum_{i=1}^{p}\hat{\mathrm{var}}(s_{i})-\hat{\mathrm{cov}}(t_{i},s_{i})-\hat{\mathrm{Bias}}(s_{i})(t_{i}-s_{i})}{\sum_{i=1}^{p}(t_{i}-s_{i})^{2}}.
\]
 Along with conventional target matrix ($I$) they used five other
target matrices summarised in the following Table. 

\subsection*{Table 1}

\begin{tabular}{|c|c|}
\hline 
\begin{tabular}{|c|}
\hline 
Target A: \textquotedbl{}diagonal, unit variance'' \tabularnewline
\hline 
0 estimated parameters\tabularnewline
\hline 
\hline 
$t_{ij}=\begin{cases}
1 & if\ i=j\\
0 & if\ i\neq j
\end{cases}$\tabularnewline
\hline 
$\hat{\lambda}^{*}=\frac{\sum_{i\neq j}\bar{var}(s_{ij})+\sum_{i=j}\bar{var}(s_{ii})}{\sum_{i\neq j}s_{ij}{}^{2}+\sum_{i}(s_{ii}-1)^{2}}$\tabularnewline
\hline 
\end{tabular} & %
\begin{tabular}{|c|}
\hline 
Target B: ``diagonal, common variance'' \tabularnewline
\hline 
1 estimated parameter: $v$\tabularnewline
\hline 
\hline 
$t_{ij}=\begin{cases}
v=avg(s_{ii}) & if\ i=j\\
0 & if\ i\neq j
\end{cases}$\tabularnewline
\hline 
$\hat{\lambda}^{*}=\frac{\sum_{i\neq j}\bar{var}(s_{ij})+\sum_{i}\bar{var}(s_{ii})}{\sum_{i\neq j}s_{ij}^{2}+\sum_{i}(s_{ij}-v)^{2}}$\tabularnewline
\hline 
\end{tabular}\tabularnewline
\hline 
\hline 
\begin{tabular}{|c|}
\hline 
Target C: \textquotedbl{}common (co)variance'' \tabularnewline
\hline 
2 estimated parameters: $v,\!c$ \tabularnewline
\hline 
\hline 
$t_{ij}=\begin{cases}
v=avg(s_{ii}) & if\ i=j\\
c=avg(s_{ij}) & if\ i\neq j
\end{cases}$\tabularnewline
\hline 
$\hat{\lambda}^{*}=\frac{\sum_{i\neq j}\bar{var}(s_{ij})+\sum_{i=j}\bar{var}(s_{ii})}{\sum_{i\neq j}(s_{ij}-c)^{2}+\sum_{i}(s_{ij}-v)^{2}}$\tabularnewline
\hline 
\end{tabular} & %
\begin{tabular}{|c|}
\hline 
Target D: \textquotedbl{}diagonal, unequal variance '' \tabularnewline
\hline 
$p$ estimated parameters: $s_{ii}$\tabularnewline
\hline 
\hline 
$t_{ij}=\begin{cases}
v=s_{ii} & if\ i=j\\
0 & if\ i\neq j
\end{cases}$\tabularnewline
\hline 
$\hat{\lambda}^{*}=\frac{\sum_{i\neq j}\bar{var}(s_{ij})}{\sum_{i\neq j}s_{ij}^{2}}$\tabularnewline
\hline 
\end{tabular}\tabularnewline
\hline 
\begin{tabular}{|c|}
\hline 
Target E: ``perfect positive correlation'' \tabularnewline
\hline 
$p$ estimated parameters: $s_{ij}$\tabularnewline
\hline 
\hline 
$t_{ij}=\begin{cases}
s_{ii} & if\ i=j\\
\sqrt{s_{ij}s_{ji}} & if\ i\neq j
\end{cases}$\tabularnewline
\hline 
$f_{ij}=\frac{1}{2}\{\sqrt{\frac{s_{jj}}{s_{ii}}}\hat{Cov}(s_{ii},s_{ij})+\sqrt{\frac{s_{ii}}{s_{jj}}}\hat{Cov}(s_{jj},s_{ij})\}$\tabularnewline
\hline 
$\hat{\lambda}^{*}=\frac{\sum_{i\neq j}\bar{Var}(s_{ij})-f_{ij}}{\sum_{i\neq j}(s_{ij}-\sqrt{s_{ii}s_{jj}})^{2}}$\tabularnewline
\hline 
\end{tabular} & %
\begin{tabular}{|c|}
\hline 
Target F: ``constant correlation'' \tabularnewline
\hline 
$p+1$ estimated parameters, $s_{ii},\!\bar{r}$\tabularnewline
\hline 
\hline 
$t_{ij}=\begin{cases}
s_{ii} & if\ i=j\\
\bar{r}\sqrt{s_{ij}s_{ji}} & if\ i\neq j
\end{cases}$\tabularnewline
\hline 
$f_{ij}=\frac{1}{2}\{\sqrt{\frac{s_{jj}}{s_{ii}}}\hat{Cov}(s_{ii},s_{ij})+\sqrt{\frac{s_{ii}}{s_{jj}}}\hat{Cov}(s_{jj},s_{ij})\}$\tabularnewline
\hline 
$\hat{\lambda}^{*}=\frac{\sum_{i\neq j}\bar{Var}(s_{ij})-\bar{r}f_{ij}}{\sum_{i\neq j}(s_{ij}-\bar{r}\sqrt{s_{ii}s_{jj}})^{2}}$\tabularnewline
\hline 
\end{tabular}\tabularnewline
\hline 
\end{tabular}

\subsection{Element-wise Regularization}

Under the assumption of sparsity, some element-wise regularization
methods can be used. In contrast to \cite{ledoit2004} type of estimator,
where only the eigenvalues were shrunk, here both eigenvalues and
vectors are regularised. We will first discuss popular methods like
Banding and Tapering which assume some order between the variables
and as a result, the estimator is not invariant under permutation
of variables. So this is useful for time-dependent data. 

\subsubsection{Banding}

The idea behind banding is that the variables are ordered in such
a way that elements of the covariance matrix, further away from the
main diagonal, are negligible. An $l-$banded covariance matrix is
defined as $B(S_{l})=[s_{ij}I(|i-j|\leq l)]$, where $S=[s_{ij}]$
is the $p\times p$ sample covariance matrix and $l$ $(\leq p)$
is band length, determined through cross validation. One can question-
which kind of population covariance matrix can be well approximated
by Banded sample covariance matrix. Intuitively, such a matrix should
have decaying entries as one moves away from the main diagonal. \cite{bickel2008covariance}
showed that the population covariance can be well approximated uniformly
over the following class of matrices: $\{\Sigma:\ \mathrm{max}_{j}\sum_{i}|\sigma_{ij}|\!I(i-j\geq k)\leq C.k^{-\alpha},\mathrm{and}\ 0<\epsilon\leq\lambda_{\mathrm{min}}(\Sigma)<\lambda_{\mathrm{max}}(\Sigma)\leq\epsilon^{-1}\}$,
where $C$ is a constant and $\alpha$ captures the rate of decay
of the entries $\sigma_{ij}$ as $i$ goes away from $j$. Although
$p$ is large, if $\mathrm{log}(p)$ is very small compared to $n$,
that is, $\frac{\mathrm{log}(p)}{n}\rightarrow0$, then such a $\Sigma$
can be well-approximated by accurately chosen band length and the
error in approximation depends on $\mathrm{log}(p)/n$ and $\alpha$.
Same result holds also for the precision matrix. Banded covariance
estimation procedure does not guarantee positive definiteness. 

\subsubsection{Tapering}

Tapering the covariance matrix is another possible way and it can
preserve positive definiteness. $T(S)=S\circ T$ is a tapered estimator
where $S$ is the sample covariance matrix, $T$ is the tapering matrix
and `$\circ$' denotes the Hadamard product (element-wise product).
Properties of Hadamard product suggest that $T(S)$ is positive definite
if $T$ is so. The banded covariance matrix is a special case of this
with $T=((1_{[|i-j|\leq l}))$, which is not positive definite. 

\subsubsection{Thresholding}

The most widely applicable element-wise regularization method is defined
through Thresholding Operator. The regularized estimator is $T_{\lambda}(S)=((s_{ij}I(s_{ij})>\lambda)\thinspace))$,
where $S=((s_{ij}))$ is the sample covariance matrix and $\lambda>0$
is the threshold parameter. $\lambda$ can be determined through cross
validation. Although it is much simpler than other methods, like penalized
lasso, it has one problem. The estimator preserves symmetry but not
positive definiteness. With Gaussian assumption, consistency of this
estimator can be shown uniformly over a class $\{\Sigma:\ \sigma_{ii}\leq C,\ \sum_{j=1}^{p}|\sigma_{ij}|^{q}\leq s_{0}(p),\ \forall\thinspace i\}$
with $0\leq q\leq1$, $\nicefrac{\mathrm{log}(p)}{n}=o(1)$ and $\lambda=M\sqrt{\frac{log(p)}{n}}$for
sufficiently large $M$ \cite{bickel2008covariance}. For $q=0$ the
condition $\sum_{j=1}^{p}|\sigma_{ij}|^{q}\leq s_{0}(p)$ reduces
to $\sum_{j=1}^{p}I(\sigma_{ij}\neq0)\leq s_{0}(p).$ The rate of
convergence is dependent on the dimension ($p$), sample size ($n$)
and $s_{0}$, the determining factor of the number of nonzero elements
in $\Sigma$. Similar result can be shown for precision matrix. For
non-Gaussian case, we need some moment conditions in order to achieve
consistency result \cite{bickel2008covariance}.

The result goes through for a larger class of thresholding operators.
One such is called generalized thresholding operators with the following
three properties: 
\begin{enumerate}
\item $|s_{\lambda}(x)|\leq|x|$ (shrinkage)
\item $s_{\lambda}(x)=0$ for $|x|\leq\lambda$(thresholding)
\item $|s_{\lambda}(x)-x|\leq\lambda$(constraint on amount of shrinkage) 
\end{enumerate}
Apart from being consistent under suitable conditions discussed earlier,
if variance of each variable is bounded then this operator is also
``sparsistent'' i.e. able to identify true zero entries of population
covariance matrix with probability tending to one.

For both thresholding and generalized thresholding operators $\lambda$
is fixed for all entries of the matrix. An adaptive threshold estimator
can be developed \cite{cai2011adaptive} to have different parameters
for different entries where 
\[
\lambda_{ij}\propto\sqrt{\frac{log(p)}{n}\hat{var}(Y_{i}-\mu_{i})(Y_{j}-\mu_{j})}
\]

\subsection{Approximate Factor Model}

Sometimes the assumption of sparsity is too much to demand. For such
situations estimation methods of a larger class of covariance matrices
is required. A simple extension is possible to the class of matrices
that can be decomposed into sum of low rank and sparse matrix: $\Sigma=FF^{T}+\Psi$,
where $F$ is low rank and $\Psi$ is sparse matrix. Due to similarity
with Factor model where $\Psi$ is diagonal, this model is called
approximate factor model. To estimate $\Sigma$, one can decompose
$S$ similarly as, $S=\sum_{i=1}^{q}\hat{\lambda}_{i}\hat{e}_{i}\hat{e}_{i}^{T}+R$,
where the first part involves the first $q$ principal components
and the second part is residual. As $R$ is sparse we can now use
thresholding/adaptive thresholding operators to estimate it \cite{chamberlain1982arbitrage}. 

\subsection{Positive Definiteness}

Sometimes positive definiteness of the estimator is required in order
to be used in classification or covariance regularised regression.
As we have discussed thresholding estimators do not guarantee positive
definite estimator. In the following section we will describe a couple
of methods to achieve that. One possible way is to replace the negative
eigenvalues in eigen decomposition of $\hat{\Sigma}$ by zero. But
this manipulation destroys the sparse nature of the covariance matrix.
An alternative way is necessary that will ensure sparsity and at the
same time will produce positive definite output. Let us denote sample
correlation matrix by $R$ matrix $M\succ0$ if it is symmetric and
positive definite ($M\succeq0$ for positive semi definite) and $M_{j,-j}=M_{-j,j}=j$-th
column of symmetric matrix $M$ with it's $j$-th element removed.
$M_{-j,-j}=$matrix formed after removing $j$-th column and $j$-th
row of $M$ and $M^{+}$ is the diagonal matrix with the same diagonal
elements as $M$. Define $M^{-}=M-M^{+}$. Then a desirable positive
definite estimator is 
\[
\hat{\Sigma}_{\lambda}=(S^{+})^{\frac{1}{2}}\hat{\Theta}_{\lambda}(S^{+})^{\frac{1}{2}}
\]
 where $S^{+}=diag(S)$ and estimated correlation matrix is 
\[
\hat{\Theta}=\mathrm{argmin}_{\Theta\succ0}{\|\Theta-R\|_{F}^{2}/2-\tau log|\Theta|+\lambda|\Theta_{-}|}
\]
 with $\lambda$and $\tau>0$ respectively being tuning parameter
and a fixed small value. The log-determinant term in the optimization
function ensures positive definiteness. Regularizing the correlation
matrix leads to faster convergence rate bound and scale invariance
of the estimator. Under suitable and reasonable conditions this estimator
is consistent\cite{rothman2012positive}. For fast computation the
following algorithm has been developed.
\begin{itemize}
\item Input $Q$- a symmetric matrix with positive diagonals, $\lambda$,
$\tau$ and initialise $(\Sigma_{0},\thinspace\Omega_{0})$ with $\Omega_{0}>0$.
Follow steps $1-3$ for $j=1,2,\ldots,p$ and repeat till convergence.

Step1: $\sigma_{jj}^{(k+1)}=q_{jj}+\tau\omega_{jj}^{(k)}$ and solve
the lasso penalized regression: 
\[
\Sigma_{j,-j}^{(k+1)}=\mathrm{argmin}_{\beta}{\frac{1}{2}\beta^{T}(I+\frac{\tau}{\sigma_{jj}^{k+1}}\Omega_{-j,-j}^{(k)})\beta-\beta^{T}Q_{-j,j}+\lambda\|\beta\|_{1}}
\]

Step2: $\Omega_{j,-j}^{(k+1)}=-\Omega_{-j,-j}^{(k)}\Sigma_{j,-j}^{(k+1)}/\sigma_{jj}^{(k+1)}.$

Step3: Compute $\omega_{jj}^{(k+1)}=(1-\Sigma_{j,-j}^{(k+1)}\Omega_{j,-j}^{(k+1)})/\sigma_{jj}^{(k+1)}.$
\end{itemize}
\noindent An alternative estimator has been proposed based on \textit{alternating
direction method} \cite{xue2012positive}. If we want a positive semi
definite matrix then the usual objective function along with $l_{1}$
penalty term should be optimized with an additional constraint for
positive semi-definiteness: 
\[
\Sigma^{+}=\mathrm{argmin}_{\Sigma\succeq0}{\|\Sigma-S\|_{F}^{2}/2+\lambda|\Sigma|_{1}}.
\]
 For positive definite matrix we can replace the constraint $\Sigma\succeq0$
with $\Sigma\succ\epsilon I$ for very small $\epsilon>0$. Introducing
a new variable $\Theta$, we can write the same as 
\[
(\hat{\Theta}_{+},\hat{\Sigma}_{+})=\mathrm{argmin}_{\Theta,\Sigma}{\|\Sigma-S\|_{F}^{2}/2+\lambda|\Sigma|_{1}:\,\Sigma=\Theta,\,\Theta\succeq\epsilon I}.
\]
 Now it is enough to minimize its augmented Lagrangian function for
some given penalty parameter $\mu$: 
\[
L(\Theta,\thinspace\Sigma;\thinspace\Lambda)=\|\Sigma-S\|_{F}^{2}/2+\lambda|\Sigma|_{1}-<\Lambda,\thinspace\Theta-\Sigma>+\nicefrac{\|\Theta-\Sigma\|_{F}^{2}}{2\mu}
\]
 , where $\Lambda$ is the Lagrange multiplier. This can be achieved
through the following algorithm ($\textbf{S }$ being Soft Thresholding
Operator):
\begin{itemize}
\item Input $\mu$, $\Sigma^{0}$, $\Lambda^{0}$. 
\item Iterative alternating direction augmented Lagrangian step: for the
$i$-th iteration: 
\begin{enumerate}
\item Solve $\Theta^{i+1}=(\Sigma^{^{i}}+\mu\Lambda^{i})_{+}$
\item Solve $\Sigma^{i+1}=\{\textbf{S}(\mu(S-\Lambda^{i})+\Theta^{i+1};\mu\lambda)\}/(1+\mu)$
\item Update $\Lambda^{i+1}=\Lambda^{i}-(\Theta^{i+1}-\Sigma^{i+1})/\mu$ 
\end{enumerate}
\item Repeat the above cycle till convergence. 
\end{itemize}

\section{Precision Matrix Estimator}

In some situations instead of covariance matrix, the precision matrix
($\Sigma^{-1}$) needs to be calculated. One example of such situation
is financial network model using partial correlation coefficients
because the sample estimate of partial correlation between two nodes
is $\hat{\rho}_{ij}=-\hat{\omega}_{ij}/\sqrt{\hat{\omega}_{ii}\hat{\omega}_{jj}}$,
where $\hat{\omega}_{ij}=(\hat{\Sigma}^{-1})_{ij}$. Of course $\hat{\Sigma}^{-1}$can
be calculated from $\hat{\Sigma}$ but that inversion involves $O(p^{3})$
operations. For high dimensional data it is computationally expensive.
On the other hand, if it is reasonable to assume sparsity of the precision
matrix, that is, most of the off-diagonal elements of the precision
matrix are zeros, then we can directly estimate the precision matrix.
Although the correlation for most of the pairs of financial institutions
would not be zero, the partial correlations can be. So this assumption
of sparsity would not be a departure from reality in many practical
situations. In such cases starting from a fully connected graph we
can proceed in a \textit{backward stepwise} fashion, by removing the
least significant edges. Instead of such sequential testing procedure,
some multiple testing strategy, for example, controlling for false
discovery rate, can also be adopted. We discuss this in detail in
section 5. After determining which off-diagonal entries of precision
matrix are zeros (by either sequential or multiple testing procedure),
maximum likelihood estimates of nonzero entries can be found by solving
a convex optimization problem: maximizing the concentrated likelihood
subject to the constraint that a subset of entries of precision matrix
equal to zero \cite{dempster1972covariance} \cite{pourahmadi2013}.

Alternatively, under Gaussian assumption a penalized likelihood approach
can be employed. If $Y_{1},...,Y_{p}\sim N_{p}(0,\Sigma)$, the likelihood
function is 
\[
L(\Sigma)=\frac{1}{(2\pi)^{np/2}|\Sigma|^{n/2}}exp(-\frac{1}{2}\sum_{i=1}^{n}Y_{i}'\Sigma^{-1}Y_{i}).
\]
 The penalized likelihood $l(\Sigma^{-1})=log|\Sigma^{-1}|-tr(S\Sigma^{-1})-\lambda\|\Sigma^{-1}\|_{1}$,
with penalty parameter $\lambda>0,$ can be used to obtain a sparse
solution \cite{yuan2007model}. The fastest algorithm to obtain the
solution is called graphical lasso \cite{friedman2008sparse}, described
as follows: 
\begin{enumerate}
\item Denote $\Theta=\Sigma^{-1}$. Start with a matrix $W$ that can be
used as a proxy of $\Sigma$. The choice recommended in Friedman et.
al is $W=S+\lambda I$.
\item Repeat till convergence for $j=1,2,\ldots,p$: 
\begin{enumerate}
\item Partition matrix $W$ in two parts, $j$th row and column, and the
matrix $W_{11}-$composed by the remaining elements. After eliminating
the $jj$th element, the remaining part of $j\text{th}$ column ($p-1$dimensional)
is denoted as $w_{12}$ and similarly the row is denoted as $w_{21}$.
Similarly, define $S_{11},$ $s_{12}$, $s_{21}$, $s_{22}$ for $S$
matrix. (For $j=p$, the partition will look like: $W=\left(\begin{array}{cc}
W_{11} & w_{12}\\
w_{21} & w_{22}
\end{array}\right)$ and $S=\left(\begin{array}{cc}
S_{11} & s_{12}\\
s_{21} & s_{22}
\end{array}\right)$).
\item Solve the estimating equations 
\[
W_{11}\beta-s_{12}+\lambda.Sign(\beta)=0,
\]

using cyclical coordinate-descent algorithm to obtain $\hat{\beta}$.
\item Update $w_{12}=W_{11}\hat{\beta}$. 
\end{enumerate}
\item In the final cycle, for each $j$, solve for $\hat{\Theta}_{12}=-\hat{\beta}\hat{\Theta}_{22}$,
with $\hat{\Theta}_{22}^{-1}=w_{22}-w_{12}'\hat{\beta}$. Stacking
up $(\hat{\Theta}_{12},\hat{\Theta}_{21})$ will give the $j$th column
of $\Theta$. 
\end{enumerate}
Figure 4 shows undirected graph from Cell-signalling data obtained
through graphical lasso with different penalty parameters \cite{friedman2008sparse}. 

\begin{figure}
\caption{Resulting networks given by graphical lasso algorithm for different
values of penalty parameter lambda \cite{friedman2008sparse} }
\label{fig4} \includegraphics[scale=0.5]{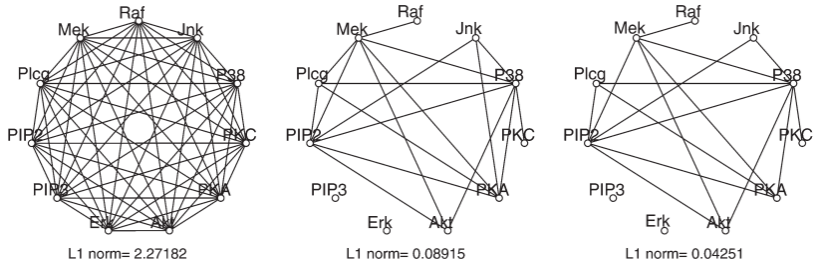} 
\end{figure}

\section{Multiple Hypothesis Testing Problem and False Discovery Rate}

We can encounter large scale hypothesis testing problems in many practical
situations. For example, in Section 4, we discussed to remove edge
from a fully connected graph, we need to perform $p(p-1)/2$ testing
problems- $H_{ij}:\!\rho_{ij}=0$ vs $K_{ij}:\!\rho_{ij}\neq0$. A
detailed review can be found in \cite{drton2007multiple}.

Suppose we have $N$ independent hypothesis $H_{1},\thinspace H_{2},...,\thinspace H_{N}$
to test. In such situations it is important to control not only the
type I error of individual hypothesis tests but also the overall (or
combined) error rate. It is due to the fact that the probability of
atleast one true hypothesis would be rejected becomes large: $1-(1-\alpha)^{N}$,
where $\alpha$ being the level of significance, generally taken as
0.05 or 0.01. The conventional way to resolve this problem is by controlling
the familywise error rate (FWER)- $P(\cup_{i=1}^{N}H_{0i}\text{ is rejected when it is true\ensuremath{)}}$.
One example of such is Bonferroni correction. The problem with this
procedure is that it is overly conservative and as a consequence the
power of the test will be small. A much more liberal and efficient
method for high dimension has been proposed by Benjamini and Hochberg
\cite{benjamini1995controlling}. In figure 5, out of $N$ hypothesis
tests in $N_{0}$ cases null hypothesis is true and in $N_{1}$cases
null hypothesis is false. According the decision rule, in $R$ out
of $N$ cases null hypothesis is rejected. Clearly $R$ is observed
but $N_{0}$ or $N_{1}$are not. The following algorithm controls
the expected false discovery proportion: 
\begin{enumerate}
\item The test statistic of $H_{1},\thinspace H_{2},...,\thinspace H_{N}$
yield $p$values $p_{1},.....,p_{N}$ .
\item Order the $p$ values $p_{(1)},.....,p_{(N)}$.
\item Rank the hypothesis $H_{1},\thinspace H_{2},...,\thinspace H_{N}$
according to the $p$values
\item Find largest $j$, say $j^{*}$, such that $p_{j}\leq\frac{j}{N}\alpha$
\item Reject the top $j^{*}$tests as significant.
\end{enumerate}
It can be shown that if the $p$ values are independent of each other
then the rule based on the algorithm controls the expected false discovery
proportion by $\alpha$, more precisely, $E(\nicefrac{a}{R})\leq\frac{N_{0}}{N}\alpha\leq\alpha$. 

\begin{figure}

\includegraphics[scale=0.5]{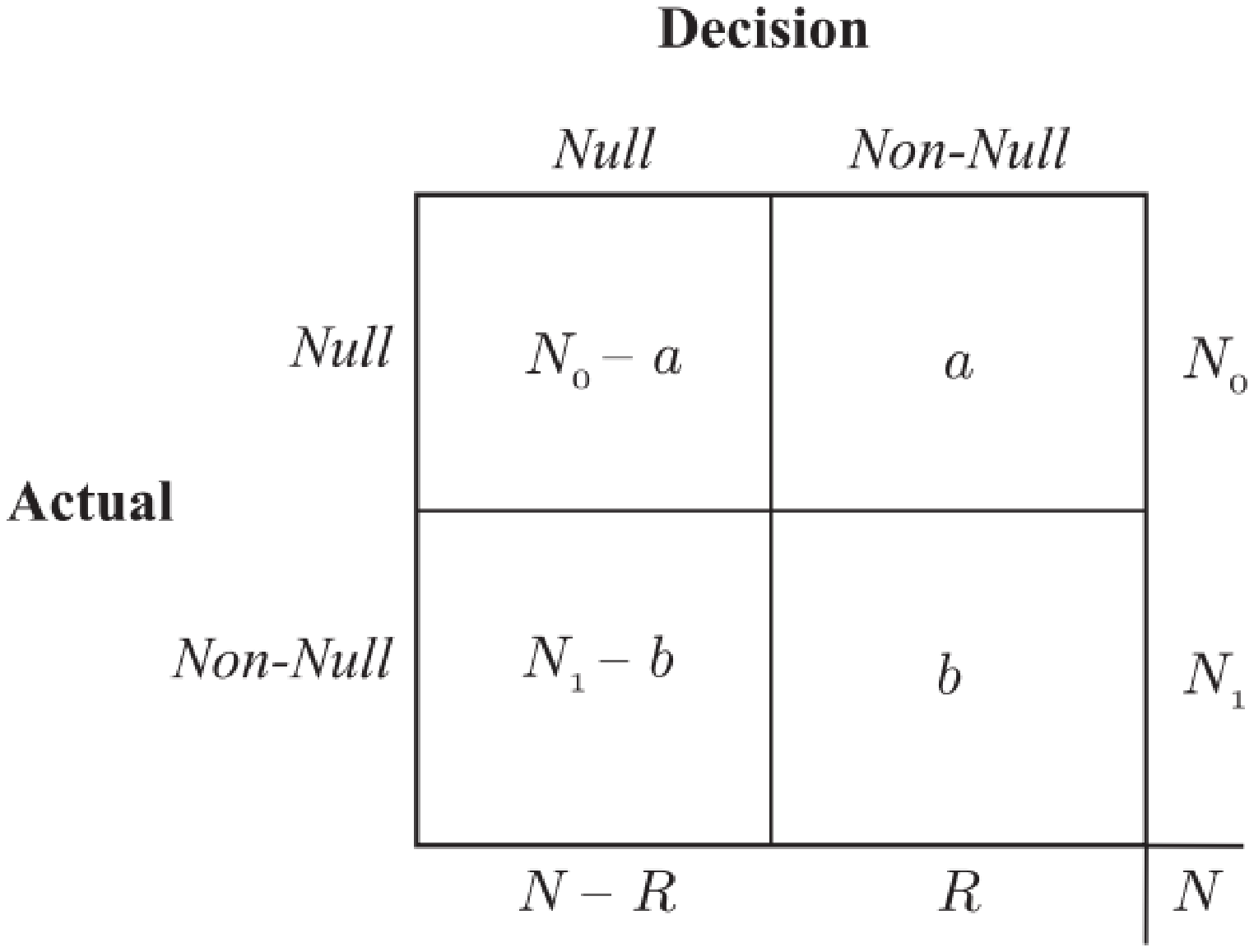}\caption{False discovery rate $\nicefrac{a}{R}$ \cite{efron2012large}}

\end{figure}

\section{High Dimensional Regression }

In financial econometrics one can often encounter multiple regression
analysis problem. A large number of predictors implies large number
of parameters to be estimated which reduces the degrees of freedom.
As a result prediction error will be increased. So in high dimensional
regression regularization is an essential tool. 

In this section we will breifly discuss the multivariate regression
problem with $q$ responses and $p$ predictors, which requires estimation
of $pq$ parameters in the regression coefficient matrix. Suppose
the matrix of regressors, responses and coefficient matrix are $X$,
$Y$ and $B$ respectively. As we know $\hat{B}_{OLS}=(X'X)^{-1}X'Y$
(under multivariate mormality, this is also the maximum likelihood
estimator) with $pq$ parameters. Estimated covariance matrix (with
$q(q+1)/2$ parameters) of $Y$ is $\hat{\Sigma}=\frac{1}{n}(Y-X\hat{B})'(Y-X\hat{B})$.
When $p$ and $q$ are large then both these estimators exhibits poor
statistical properties. So here again, shrinkage and regularization
of $\hat{B}$ would help to obtain a better estimator. It can be achieved
through Reduced Rank Regression which attempts to solve a constrained
least square problem: 
\[
\hat{B}_{r}=\underset{B:\ \mathrm{rank}(B)=r\leq\mathrm{min}(p,q)}{\mathrm{argmin}}tr[(Y-XB)'(Y-XB)]
\]
 The solution of this constrained optimization is $\hat{B}_{r}=(X'X)^{-1}X'YHH'$
where $H=(h_{1},....,h_{r})$ with $h_{k}$ being normalized eigenvector
corresponding to the $k$th largest eigenvalue of the matrix $Y'X(X'X)^{-1}X'Y$.
Choice of $r$ is important because this is the parameter that balances
between bias and variance of prediction. 

Alternatively, a regularized estimator can be obtained by introducing
a nonnegative penalty function in the optimization problem: 
\[
\hat{B}=\mathrm{argmin_{B}}\{tr[(Y-XB)'(Y-XB)]+\lambda C(B)\}
\]
 when $C$ is a scalar function and $\lambda$ is nonnegative quantity.
Most common choices of $C(B)$ are $l_{p}$ norms. $C(B)=\sum_{j,k}|b_{jk}|$
leads to lasso estimate where as $C(B)=\sum_{j,k}b_{jk}^{2}$ amount
to ridge regression. $C(B)=\alpha\sum_{j,k}|b_{jk}|+\nicefrac{(1-\alpha)}{2}\sum_{j,k}b_{jk}^{2}$
for $\alpha\in(0,1)$ and $C(B)=\sum_{j,k}|b_{jk}|^{\gamma}$ for
$\gamma\in[1,2]$ are called elastic net and bridge regression respectively.
Grouped lasso with $C(B)=\sum_{i=1}^{p}(b_{j1}^{2}+...+b_{jq}^{2})^{0,5}$
imposes groupwise penalty on the rows of $B$, which may lead to exclusion
of some predictors for all the responses. 

All the above mentioned methods regularize the matrix $B$ only while
leaving $\Sigma$aside. Although a little more complicated, it is
sometimes appropriate to regularize both $B$ and $\Sigma^{-1}$.
One way to do that is to adding separate lasso penalty on $B$ and
$\Sigma^{-1}$in the negative log likelihood: 
\[
l(B,\Sigma)=\mathrm{tr}[\frac{1}{n}(Y-XB)\Sigma^{-1}(Y-XB)]-\mathrm{log}|\Sigma^{-1}|+\lambda_{1}\sum_{j'\neq j}|\omega_{j'j}|+\lambda_{2}\sum_{j,k}|b_{j,k}|
\]

where $\lambda_{1}$ and $\lambda_{2}$ are as usual tuning parameter,
$B=((b_{jk}))$ and $\Sigma^{-1}=\Omega=((\omega_{j',j}))$. This
optimization problem is not convex but biconvex. Note that solving
the above mentioned optimization problem for $\Omega$ with $B$ fixed
at $B_{0}$ reduced to the optimization problem:
\[
\hat{\Omega}(B_{0})=\mathrm{argmin_{\Omega}}\{tr(\hat{\Sigma}\Omega)-log|\Omega|+\lambda_{1}\sum_{i\neq j}|\omega_{ij}|\}
\]
where $\hat{\Sigma}=\frac{1}{n}(Y-XB_{0})'(Y-XB_{0})$. If we fix
$\Omega$ at a nonnegative definite $\Omega_{0}$ it will lead to
\[
\hat{B}(\Omega_{0})=\mathrm{argmin_{B}}\{tr[\frac{1}{n}(Y-XB_{0})\Omega(Y-XB_{0})']+\lambda_{2}\sum_{j,k}|b_{j,k}|\}
\]
It can be shown that the original problem can be solved by using the
following algorithm prescribed by \cite{rothman2010sparse}- 
\begin{itemize}
\item Fix $\lambda_{1}$ and $\lambda_{2}$, initialize $\hat{B}^{(0)}=0$
and $\hat{\Omega}^{(0)}=\hat{\Omega}(\hat{B}^{(0)})$.
\begin{itemize}
\item Step 1: Compute $\hat{B}^{(m+1)}=\hat{B}(\hat{\Omega}^{(m)})$ by
solving 
\[
\mathrm{argmin_{B}}\{tr(\frac{1}{n}(Y-XB)\Omega(Y-XB)+\lambda_{2}\sum_{j}\sum_{k}|b_{jk}|\}
\]
 by coordinate descent algorithm.
\item Step 2: Compute $\hat{\Omega}^{(m+1)}=\hat{\Omega}(\hat{B}^{(m+1)})$
by solving 
\[
\mathrm{argmin_{\Omega}}\{tr(\hat{\Sigma}\Omega)-log|\Omega|+\lambda_{1}\sum_{i\neq j}|\omega_{ij}|\}
\]
 by Graphical lasso algorithm. 
\item Step 3: If $\sum_{i,j}|\hat{b}_{ij}^{(m+1)}-\hat{b}_{ij}^{(m)}|<\epsilon\sum_{i,j}\hat{b}_{ij}^{R}$
where $((\hat{b}_{ij}^{R}))$ is the Ridge estimator of $B$.
\end{itemize}
\end{itemize}

\section{Principal Components}

In many high dimensional studies estimates of principal component
loadings are inconsistent and the eigenvectors consists of too many
entries to interpret. In such situation regularising the eigenvectors
along with eigenvalues would be preferable. So it is desirable to
have loading vectors with only a few nonzero entries. The simplest
way to achieve that is through a procedure called SCoTLASS \cite{jolliffe2003modified}.
In this approach a lasso penalty is to be imposed on the PC loadings.
So the first PC loading can be obtained by solving the optimization
problem: 
\[
\mathrm{maximize_{v}}vX'Xv\ \mathrm{subject\ to}\ ||v||_{2}^{2}\leq1,\ ||v||_{1}\leq c
\]

The next PC can be obtained by imposing extra constraint of orthogonality.
Note that this is not a minimization problem and so can be difficult
to solve. However the above problem is equivalent to the following
one:
\[
\mathrm{maximize_{u,v}}u'Xv\ \mathrm{subject\ to}\ ||v||_{1}\leq c,\ ||v||_{2}^{2}\leq1,\ ||u||_{2}^{2}\leq1
\]
The equivalence between the two can be easily verified by using Cauchy
Schwartz inequality to $u'Xv$ and noting that equality will be achieved
for $u=\frac{X'v}{||X'v||_{2}}$. The optimization problem can be
solved by the following algorithm \cite{witten2009penalized}
\begin{itemize}
\item Initialize $v$ to have $l_{2}$ norm 1. 
\item Iterate until convergence
\begin{lyxlist}{00.00.0000}
\item [{a)}] $u\leftarrow\frac{Xv}{||Xv||_{2}}$
\item [{b)}] $v\leftarrow\frac{s(X'u,\Delta)}{||s(X'u,\Delta)||_{2}}$,
where $S$ is a soft thresholding operator, and $\Delta=0$ if the
computed $v$ satisfies $||v||_{1}\leq c;$ otherwise $\Delta>0$
with $||v||_{1}=c$
\end{lyxlist}
\end{itemize}

\section{Classification}

Suppose there are $n$ independent observations of training data ($\textbf{X}_{i},\ Y_{i}$),
$i=1(1)n$, coming from an unknown distribution. Here $Y_{i}$ denotes
the class of the $i$th observation and therefore can take values
$\{1,2,3,...,K\}$ if there are $K$ classes. $\textbf{X }_{i}$ ,
generally a vector of dimension $p$, is the feature vector for the
ith observation. Given a new observation $\textbf{X}$, the task is
to determine the class, the observation belongs to. In other words
we have to determine a function from the feature space to $\{1,2,...,K\}$.
One very widely used class of classifiers is distance based classifiers.
It assigns an observation to a class $k$, if the observation is closer
to class $k$ on average compared to other classes i.e. $k=\mathrm{argmin_{i}\thinspace dist}(\textbf{X},\mu_{i})$,
where $\mu_{i}$ is the center of the feature space of class $i$
. As an example if there are two classes and the feature distribution
for the first class is $\textbf{X }\sim N(\mu_{1},\Sigma)$ and for
the second class is $\textbf{X}\sim N(\mu_{2},\Sigma)$ then under
the assumption that both the classes have equal prior probabilities,
the most widely used distance measure is called Mahalanobis's distance
\[
dist(X,\mu_{k})=\sqrt{(X-\mu_{k})\Sigma^{-1}(X-\mu_{k})},k=1,2
\]
. So class 1 is to be chosen when 
\[
\sqrt{(X-\mu_{1})\Sigma^{-1}(X-\mu_{1})}\leq\sqrt{(X-\mu_{2})\Sigma^{-1}(X-\mu_{2})}
\]

This technique is called Fisher Discriminant Analysis. For high dimensional
data Fisher Discriminant Analysis does not perform well because it
involves accurate estimation of precision matrix \cite{bickel2004some}.
In the following section we will discuss some high dimensional classification
methods. 

\subsection{Naive Bayes Classifier}

Suppose we classify the observation, with feature $\textbf{x}$, by
some predetermined function $\delta$ i.e. $\delta(\textbf{x})\in\{1,2,..,K\}$.
Now to judge the accuracy we need to consider some Loss function.
A most intuitive loss function is the zero-one loss: $L(\delta(\textbf{x}),Y)=I(\delta(\textbf{x})\neq Y$),
where $I(.)$ is the indicator function. Risk of $\delta$ is the
expected loss- $E(L(\delta(\textbf{x}),Y))=1-P(Y=\delta(\textbf{x})|\textbf{X}=\textbf{x})$.
The optimal classifier, minimizing the risk, is $g(\textbf{x})=\mathrm{arg\thinspace max_{k}}P(Y=k|\textbf{X}=\textbf{x})$.
If $\pi$ be the prior probability of an observation being in class
$k$ then by Bayes Theorem $P(Y=k|\textbf{X}=\textbf{x})=\frac{\pi(k)P(\textbf{X}=\textbf{x}|Y=k)}{\sum\pi(k)P(\textbf{X}=\textbf{x}|Y=k)}$.
So $g(\textbf{X})=\mathrm{arg\thinspace max_{k}}\pi(k)P(\textbf{X}=\textbf{x}|Y=k)$.
This is called Bayes classifier. When $\textbf{X}$ is high dimensional
$P(\textbf{X |}Y)$ is practically impossible to estimate. The Naive
Bayes Classifier works by assuming conditional independence: $P(\textbf{X}=\textbf{x}|Y=k)=\prod_{i}P(X_{i}=x_{i}|Y=k)$
where $X_{j}$ is the $j$th component of $\textbf{X}$. Naive Bayes
classifier is used in practice even when the conditional independent
assumption is not valid. In case of the previous example, under some
conditions Naive Bayes classifier outperforms Fisher Discriminant
function as long as dimensionality $p$ does not grow faster than
sample size $n$. 

\subsection{Centroid Rule and k-Nearest-Neighbour Rule}

The centroid rule classifies an observation to $k$th class if its
distance to the centroid of $k$th class is less than that to the
centroid of any other class. The merit of this method is illustrated
for $K=2$. Suppose $n_{1}$and $n_{2}$ are fixed and $p\rightarrow\infty$
and within each class observations are iid. The observation of two
classes can be denoted by $Z_{1}=(Z_{11},Z_{12},...,Z_{1p})$ and
$Z_{2}=(Z_{21},...,Z_{2p})$ respectively. With the assumption that
as $p\rightarrow\infty$, $\frac{1}{p}\sum_{i=1}^{p}var(Z_{1i})\rightarrow\sigma^{2}$,
$\frac{1}{p}\sum_{i=1}^{p}var(Z_{2i})\rightarrow\tau^{2}$ , $\sigma^{2}/n_{1}>\tau^{2}/n_{2}$
and $\frac{1}{p}\sum_{i=1}^{p}[E(Z_{1i}^{2})-E(Z_{2i}^{2})]\rightarrow\kappa^{2}$
a new observation is correctly classified with probability converging
to 1 as $p\rightarrow\infty$ if $\kappa^{2}\geq\nicefrac{\sigma^{2}}{n_{1}}-\nicefrac{\tau^{2}}{n_{2}}$\cite{hall2005geometric}.

The $k$-Nearest Neighbour rule determines the class of a new observations
by help of $k$ nearest data points from the training data. The new
observation is to be assigned to the class closest to 
\[
g(\textbf{X})=\frac{1}{k}\sum_{i:\!\textbf{X}_{i}\in N_{k}(\textbf{X})}Y_{i}
\]
 where $N_{k}(\textbf{x})$ is the set $k$ nearest points around
$\textbf{x}$. 

\subsection{Support Vector Machine}

In Bayes classifier, as we discussed, one tries to minimize $\sum_{i}I(g(X)\neq Y)$,
with respect to $g(.)$. But it is difficult to work with as indicator
function is neither smooth nor convex. So one can think of using a
convex loss function. Support vector machine (SVM) claims to resolve
that problem. Suppose for binary classification problem, $Y$ takes
-1 and 1 to denote two classes. The SVM replaces zero-one loss by
convex hinge loss $H(x)=[1-x]_{+}$ where $[u]_{+}=max\{0,u\}$, the
positive part of $u$. The SVM tries to minimize $\sum_{i}H(Y_{i}g(\textbf{X}_{i}))+\lambda J(g)$
with respect to $g$. Here $\lambda$ is a tunning parameter and $J$
is a complexity penalty of $g$. If the minimizer is $\hat{g}$ then
the SVM classifier is taken to be $sign(\hat{g})$. $J(.)$ can be
taken as $L_{2}$ penalty. 

It can be shown that the fuction minimizing $E(H(Yg(\textbf{X}))+\lambda J(g))$
is exactly $sign(P(Y=+1|\textbf{X=}\textbf{x})-\frac{1}{2})$ \cite{lin2002support}.
Infact instead of working with $P(Y|\textbf{X}=\textbf{x})$ as in
Bayes classifier SVM directly tries to estimate the decision boundary
$\{\textbf{x}:\ P(Y=1|\textbf{X}=\textbf{x})=\frac{1}{2}\}$. 

\subsection{AdaBoost}

Among the recently developed methodologies one of the most important
is Boosting. It is a method that combines a number of 'weak' classifiers
to form a powerful 'committee'. AdaBoost is the most commonly used
boosting algorithm. An interesting result of this algorithm is that
it is immuned to overfitting i.e. the test error decreases consistently
as more and more classifiers are added. Suppose we have two classes
represented as $-1$ and $1$ and denoted by $y$. We have $n$ training
data points $(x_{1},y_{1}),\ ,(x_{2},y_{2}),\ ,...,(x_{n},y_{n}).$
We want to produce a committee $F(x)=\sum_{i=1}^{M}c_{m}f_{m}(x)$
where $f_{m}$ is a weak classofier (with values either $1$ or $-1$)
that predicts better than random guess. The initial classifiers $f_{m}$
are trained on an weighted version of training sample giving more
weight to the cases that are currently misclassified. The ultimate
prediction will be based on $sign(F(x))$. This following algorithm
is called discrete AdaBoost (as the initial classifier can take only
two discrete values +1 and -1) \cite{friedman2000additive} . 

\textbf{Discrete AdaBoost algorithm}: 
\begin{enumerate}
\item Start with weights $w_{i}=\nicefrac{1}{n}$ for $i=1(1)n$
\item Repeat for $m=1(1)M$:
\begin{enumerate}
\item Fit the classifier $f_{m}(x)\in\{-1,1\}$ with weights $w_{i}$ on
training data. 
\item Compute error $e_{m}=E_{w}[I_{(y\neq f_{m}(x))}]$, where $E_{w}$
is the expectation over the training data with weights $w=(w_{1,}w_{2},...,w_{n})$
and $I(.)$ is an indicator function.
\item $c_{m}=log(\frac{1-e_{m}}{e_{m}})$
\item Set $w_{i}\leftarrow w_{i}exp[c_{m}I_{(y_{i}\neq f_{m}(x_{i}))}]$
for $i=1,2,..,n$ and then renormalize to get $\sum_{i}w_{i}=1$. 
\end{enumerate}
\item Final classifier: $sign(\sum_{m=1}^{M}c_{m}f_{m}(x))$
\end{enumerate}
The base classifier ($f_{m}(.)$) of Discrete AdaBoost algorithm is
binary. It can be generalized further to obtain a modification over
discrete AdaBoost algorithm. 

\textbf{Real AdaBoost algorithm}: 
\begin{enumerate}
\item Start with weights $w_{i}=\nicefrac{1}{n}$ for $i=1(1)n$.
\item Repeat for $m=1(1)M$:
\begin{enumerate}
\item Fit the classifier to obtain the class probability estimate $p_{m}(x)=\hat{P}_{w}(y=1|x)\in[0,1]$,
with weights $w=(w_{1},w_{2},...,w_{n})$ on the training data.
\item Set $f_{m}(x)\leftarrow\frac{1}{2}log\thinspace(p_{m}(x)/(1-p_{m}(x)))\in\mathbf{R}$. 
\item Set $w_{i}\leftarrow w_{i}exp[-y_{i}f_{m}(x_{i})]$, $i=1,2,..,n$
and renormalize so that $\sum_{i}w_{i}=1$. 
\end{enumerate}
\item Finale classifier $sign[\sum_{m}f_{m}(x)]$. 
\end{enumerate}
It can be shown that AdaBoost method of classification is equivalent
to fiiting an additive logistic regression model in a forward stagewise
manner \cite{friedman2000additive}.

\section*{Concluding remarks}

With the availability of high dimensional economic and financial data
many classical statistical methods do not perform well. We have discussed
some of the commonly encounted problems related to inference for high
dimensional financial data. In many of the approaches signicant improvement
can be achieved by bias variance trade off. Some feasible solutions
to the problems and some efficient algorithms are discussed. It is
to be noted that there are many other challeges related to high dimensional
data. Some solutions have been proposed based on simulation studies
without desired theoretical justifications. 

\bibliographystyle{plain}
\bibliography{review}

\end{document}